\documentclass[11pt,twoside]{article}
\usepackage{asp2014}

\def\be{\begin{equation}}
\def\ee{\end{equation}}

\def\deg{$^\circ$}

\def\kms{km s$^{-1}$}

\def\be{\begin{equation}}
\def\bea{\begin{eqnarray}}
\def\ee{\end{equation}}
\def\eea{\end{eqnarray}}
\def\HII{\hbox{H\,{\sc ii}}}

\def\arcmin{$^\prime$}

\def\R0{R$_{\odot}$}

\def\pccm6{pc cm$^{-6}$}

\newcommand{\Msun}{$\ensuremath\,M_\odot$}
\catcode`\@=11
\newcommand{\gsim}{${\mathrel{\mathpalette\@versim>}}$}
\newcommand{\lsim}{${\mathrel{\mathpalette\@versim<}}$}
\newcommand{\@versim}[2]{\lower 2.9truept \vbox{\baselineskip 0pt \lineskip
    0.5truept \ialign{$\m@th#1\hfil##\hfil$\crcr#2\crcr\sim\crcr}}}
\catcode`\@=12

\aspSuppressVolSlug
\resetcounters

\begin{document}

\title{Radio Recombination Line Observations Toward the Massive Star Forming Region W51 IRS1}
\author{Mishaal.~I.~Jan$^1$, D.~Anish~Roshi$^2$, M. E. Lebr\'on$^3$, E. Pacheco$^4$, T. Ghosh$^5$,\\ C. J. Salter$^{5,2}$, R. Minchin$^6$, E. D. Araya$^7$, H. G. Arce$^8$ \\ 
\affil{$^1$University of Colorado at Boulder, Boulder, Colorado 80309, USA; \email{Mishaal.Jan@colorado.edu}}
\affil{$^2$Arecibo Observatory, Arecibo, PR 00612, USA; \email{aroshi@naic.edu}}
\affil{$^3$ Department of Physical Sciences, University of Puerto Rico, R\'{i}o Piedras Campus, San Juan, PR 00931, USA}
\affil{$^4$ Department of Physics, University of Puerto Rico, R\'{i}o Piedras Campus, San Juan, PR 00931, USA}
\affil{$^5$Green Bank Observatory, P.O.Box 2, Green Bank, WV 24944, USA} 
\affil{$^6$USRA/SOFIA Science Center, NASA Ames Research Center, Moffett Field, CA 94035, USA} 
\affil{$^7$Western Illinois University, 1 University Circle, Macomb, IL 61455, USA} 
\affil{$^8$Department of Astronomy, Yale University, PO Box 208101, New Haven CT 06520, USA}
}

\paperauthor{M.~I.~Jan}{Mishaal.Jan@colorado.edu}{}{University of Colorado at Boulder}{Physics}{Boulder}{Colorado}{80309}{USA}
\paperauthor{D. Anish Roshi}{anish.roshi@gmail.com}{0000-0002-1732-5990}{Arecibo Observatory}{}{Arecibo}{PR}{00612}{USA}

\begin{abstract}
We observed radio recombination lines (RRLs) toward the W51 molecular cloud complex, one of the most active star forming regions in our Galaxy. The UV radiation from young massive stars ionizes gas surrounding them to produce \HII\ regions. 
Observations of the W51 IRS1 \HII\ region were made with the Arecibo 305 m telescope.
Of the full 1-10 GHz database, we have analyzed the observations between 4.5 and 5 GHz here. The steps involved in the analysis were: a) bandpass calibration using on-source/off-source observations; b) flux density calibration; c) removing spectral baselines due to errors in bandpass calibration and d)
Gaussian fitting of the detected lines. We detected alpha, beta and gamma transitions of hydrogen and alpha transitions of helium. 
We used the observed line parameters to 1) measure the source velocity (56.6 $\pm$ 0.3 km s$^{-1}$) with respect to
the Local Standard of Rest (LSR); 2) estimate the electron temperature (8500 $\pm$ 1800 K) of the \HII\ region and 3) derive the emission measure (5.4 $\pm$ 2.7 $\times$ 10$^{6}$ pc cm$^{-6}$) of the ionized gas.
\end{abstract}

\section{Introduction}

The giant molecular cloud complex W51 is one of the most massive and active star forming regions in our galaxy \citep[see][]{carpsan1998, getal12, lim19}. The complex was first discovered through the radio continuum emission from the embedded \HII\ regions \citep{w58}. Subsequent low-frequency ($<$ 500 MHz) continuum images of the complex made with the Arecibo 305-m Telescope indicated that the region consists of multiple thermal and non-thermal sources. The two main thermal sources are the W51A and W51B  \HII\ region complexes while the non-thermal source is the supernova remnant W51C \citep{kv67}. The W51A \HII\ region complex, located at $l$=49\deg.5, $b$=$-$0\deg.4, shows two infrared emission peaks - IRS1 and IRS2 - in the observations made by \nocite{wetal74}Wynn-Williams et al. (1974), 
which are the location of the most active and youngest ($\sim$0.8 Myr) star formation regions in W51. Both are rich in maser emission. The maser parallax measurement of W51 e2 and e8 (see Fig~\ref{fig1}) provides a distance of 5.41$^{+0.31}_{-0.28}$ kpc \citep{setal10}. The parallax distance measured toward W51 IRS2 is 5.1$^{2.9}_{-1.4}$ kpc \citep{xetal09}. The mass of W51A estimated from the 56.9 \kms\ component of CO line emission at this distance is 1.8 $\times$ 10$^{5}$ \Msun\ \citep{retal10}. Infrared observations have shown that a massive cluster with at least 20 OB stars, a plethora of young stellar objects (YSO), many of them massive young stellar objects (MYSOs), and at least one currently accreting O3 star are embedded in the W51A region \citep{getal16, betal16, lim19}. 

The embedded massive clusters in W51A ionize the material surrounding them producing the \HII\ region complex. We have observed radio recombination lines from this \HII\ region co-existing with IRS1 over the frequency range from 1 to 10 GHz with the Arecibo 305-m telescope. The data were obtained as part of a program to survey molecular and recombination lines toward W51. In this paper, we focus on the processing of data over the frequency range 4.5 to 5 GHz. We have used this data set to estimate the properties of the \HII\ region, using the recombination lines in this band, and compare the results with those available in the literature. Descriptions of the observations and data processing are given in Section~\ref{datproc}. The results are presented in Section~\ref{result}.

\section{Observation and Data Analysis}
\label{datproc}

\articlefigure{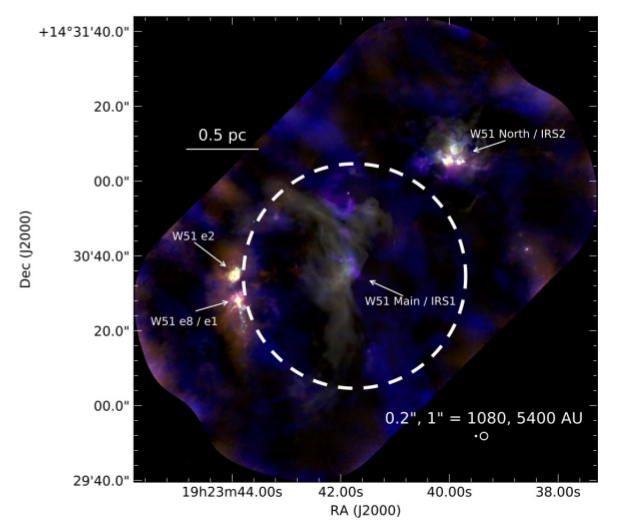}{fig1}{Composite image of W51A made from molecular line (CO in blue, CH$_3$OH in orange \& HC$_3$N in purple) and radio continuum (1.3\,mm in green \& 2.1\,cm in white) emission observed with the ALMA and JVLA \citep{g17} 
The recombination line observations toward W51 IRS1 presented in this paper were made with the Arecibo Telescope over the frequency range 4.5 to 5 GHz. The telescope beam of 1\arcmin~at 4.5 GHz is shown by the dashed circle. The dot and circle shown in the bottom-right represent roughly 0.2\arcsec\ and 2\arcsec\ respectively. 
}

The observations toward W51 IRS1 were made with the Arecibo 305-m telescope over the frequency range 1 to 10 GHz (Arecibo project code A2463). The data were collected between April 2010 and December 2014 using six different receivers to cover the full frequency range. All 14 `boxes' of the Mock Spectrometer were employed in the device's single-pixel mode. The band between 4 and 5 GHz was covered in two frequency settings, 4.0-4.5 and 4.5-5.0 GHz, with a spectral resolution of 5.25 kHz.
In this paper, we present results from the analysis of the data over the frequency range 4.5 to 5 GHz. The angular resolution of the telescope at this frequency is $\sim$ 1\arcmin.  

We observed in a standard position switched mode with 5 minutes integration on the source and similar integration at an off-source position. The data at the off-source position were used for system bandpass calibration. The \HII\ region in W51 IRS1 has strong continuum emission relative to the off-source position. This strong emission will produce a residual spectral baseline ripple in the calibrated spectrum obtained with position switched observing. The residual baseline ripple can be significantly reduced by a technique referred to as `double position switching' \citep{gs01}. To apply this technique, we observed the calibrator source 3C 390 in position switching mode with roughly the same integration time. In this paper, we present results obtained with the standard position switching mode of calibration. Results obtained with the `double position switching' technique will be presented elsewhere. 

The bandpass calibrated spectra obtained for each frequency setting were averaged separately to obtain the final spectrum. The amplitude of the average spectrum was converted to flux density in two steps. The signal from a calibrated noise diode was injected at the off-source position, and used to convert the spectral amplitude to antenna temperature. Telescope gain values (in K/Jy) provided by the observatory were used to convert the antenna temperature to Jy. A correction factor was estimated for each frequency setting using the observations of 3C 390, and these were applied in the second step to take into account small inaccuracies in the assumed gain and noise temperature calibrations. A flux density model for 3C 390 was required to obtain this correction factor. We used the measured flux densities of 3C 390 at different frequencies listed in NED\footnote{The NASA/IPAC Extragalactic Database (NED) is operated by the Jet Propulsion Laboratory, California Institute of Technology, under contract with the National Aeronautics and Space Administration.} to construct the flux density model. The model obtained is
\begin{equation}
    \textrm{log}_{10}(S) = 0.81 - 0.83\; \textrm{log}_{10}(\nu_{\textrm{GHz}}), 
\end{equation}
where $S$ is the flux density in Jy and $\nu_{\textrm{GHz}}$ is the frequency in GHz. We assume that the variability of the source over the observing period is negligible. 
Fig.~\ref{fig2} shows the uncorrected 
and corrected spectra measured toward 3C 390 and W51 IRS1. The residual peak to peak amplitude variation in the final integrated spectrum toward W51 IRS1 is about 10\%. We attribute this variation to the non-linearity in the receivers when observing a strong source. The non-linearity will introduce a change in system gain between on and off-source observations. This change in gain can be measured by performing measurements with noise injection both at the off-source and the on-source positions. Currently such measurement is not available and thus our peak-to-peak flux density variation in the observed frequency range is about 10\%.  

\articlefiguretwo{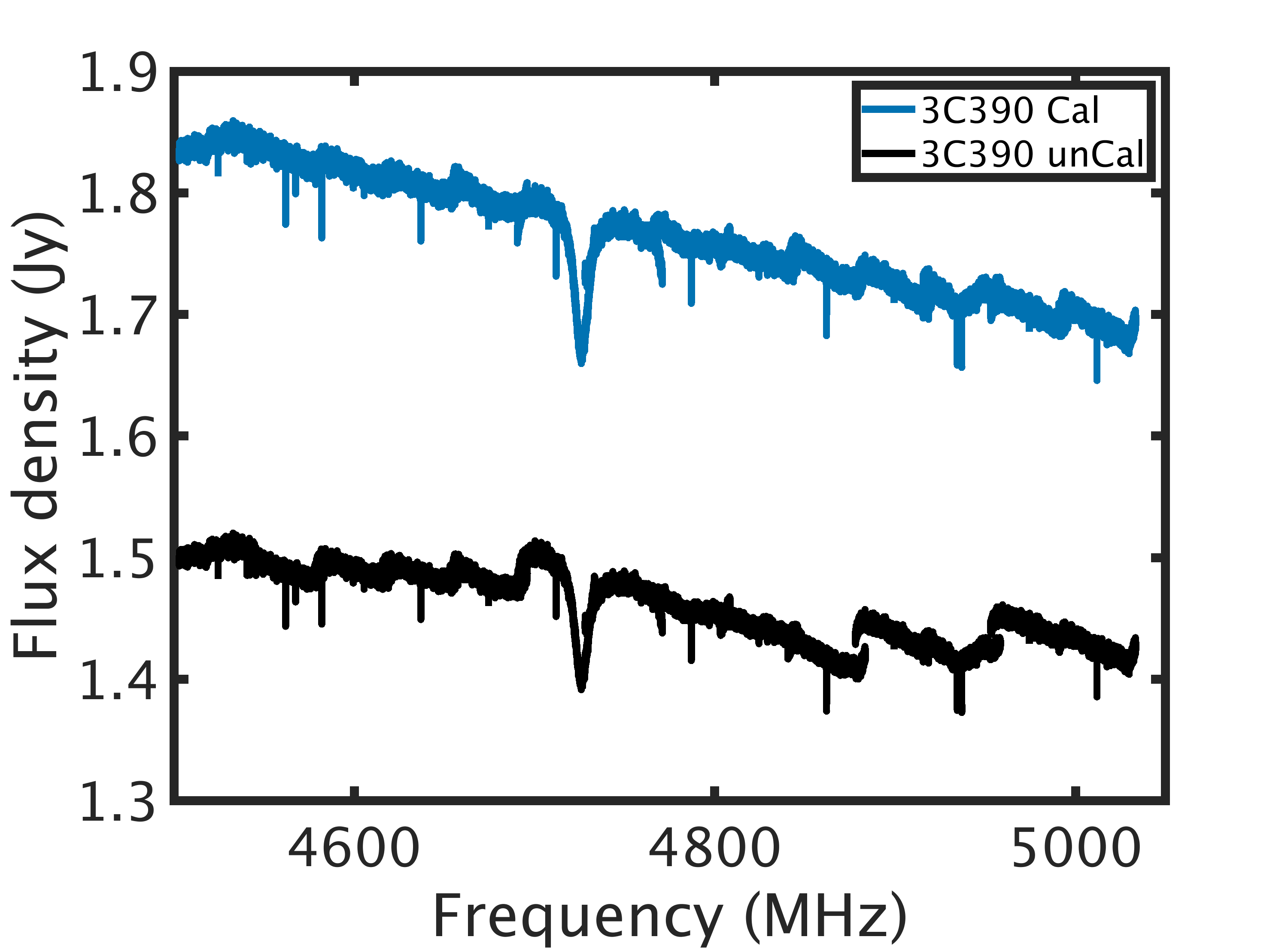}{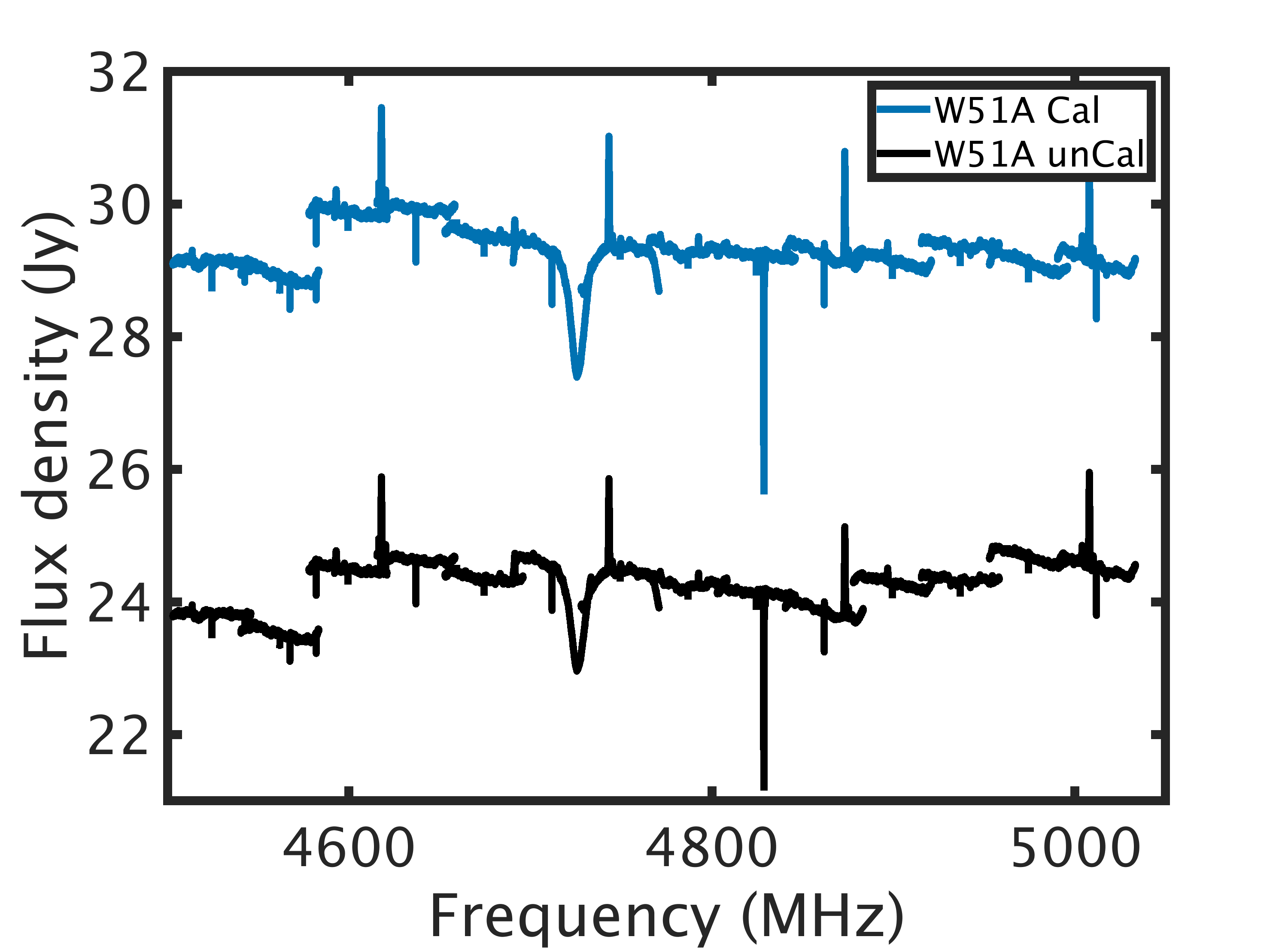}{fig2}{Corrected (shown in blue) and un-corrected (shown in black) spectra toward 3C 390 and W51A (see text).The `spectral feature' seen close to 4725 MHz is due to a resonance in the receiver system.}

The flux density corrected spectra have baseline `ripples' due to residual bandpass calibration errors which, could not be removed by a low order polynomial fit. Therefore, we first identified relatively bright recombination lines that were at least 3 times the RMS of the baseline ripple from their expected observing frequency. The spectral range adjacent to the identified lines (up to $\sim$ 50 \kms\ on either side of the half power width in the case of a single line or on either side of the `edge' lines in the case of multiple, overlapping lines) were selected for detailed baseline fitting. A polynomial of order $\le$ 3 was removed from the selected spectral range. Fig.~\ref{fig3} shows an example spectrum before and after removing a polynomial baseline. The observed frequencies of the selected spectral range were then converted to LSR (local standard of rest) velocities using the rest frequency of the identified recombination line. The line parameters were obtained by modeling the line profile as Gaussian components.  

\articlefiguretwo{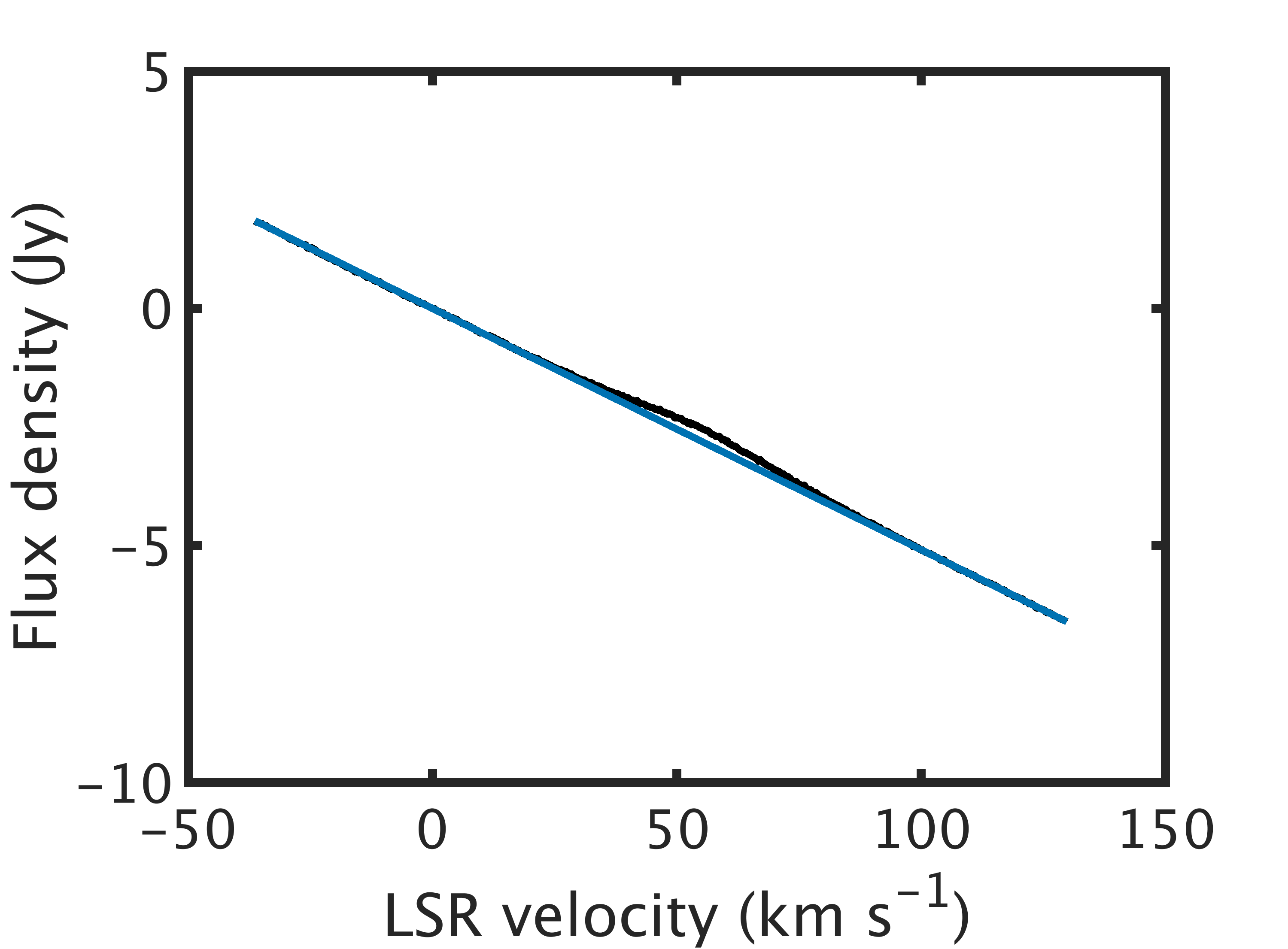}{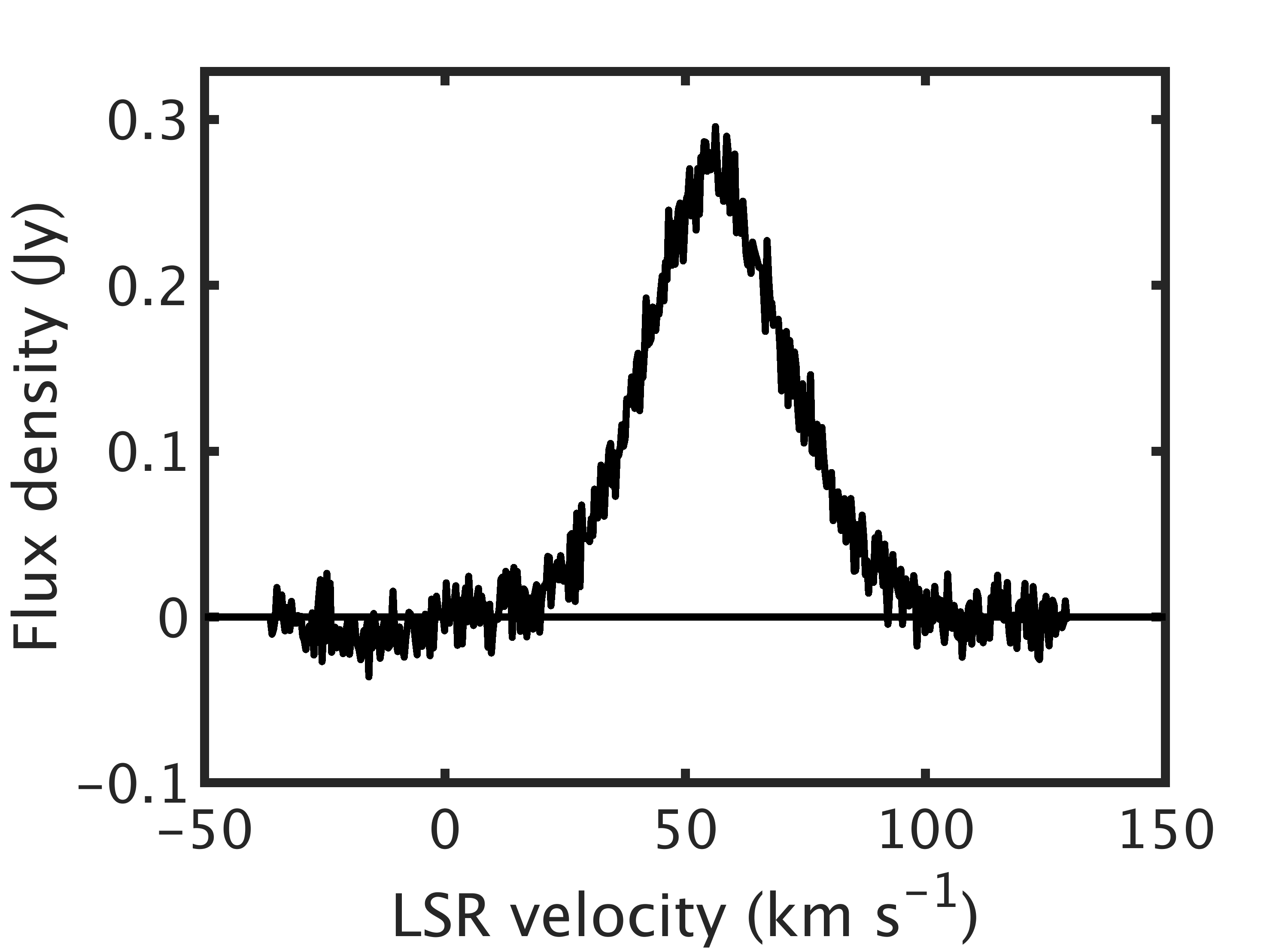}{fig3}{ \emph{Left:} The selected spectral range (black) for recombination line H140$\beta$ and the least squares polynomial fit (blue).  \emph{Right:} The spectrum after subtraction of the polynomial baseline.}

\section{Results and Discussion}
\label{result}

\articlefigurefour{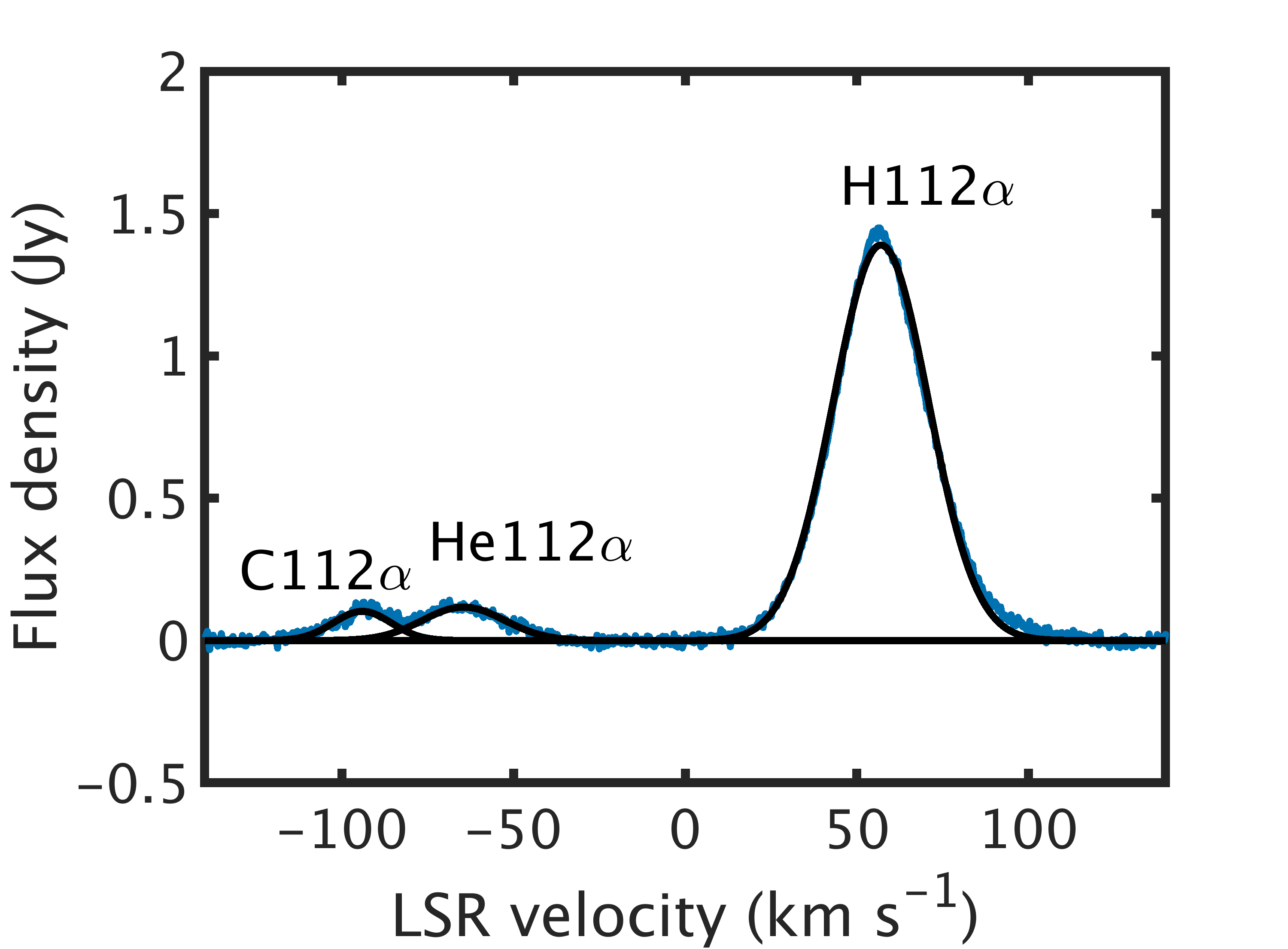}{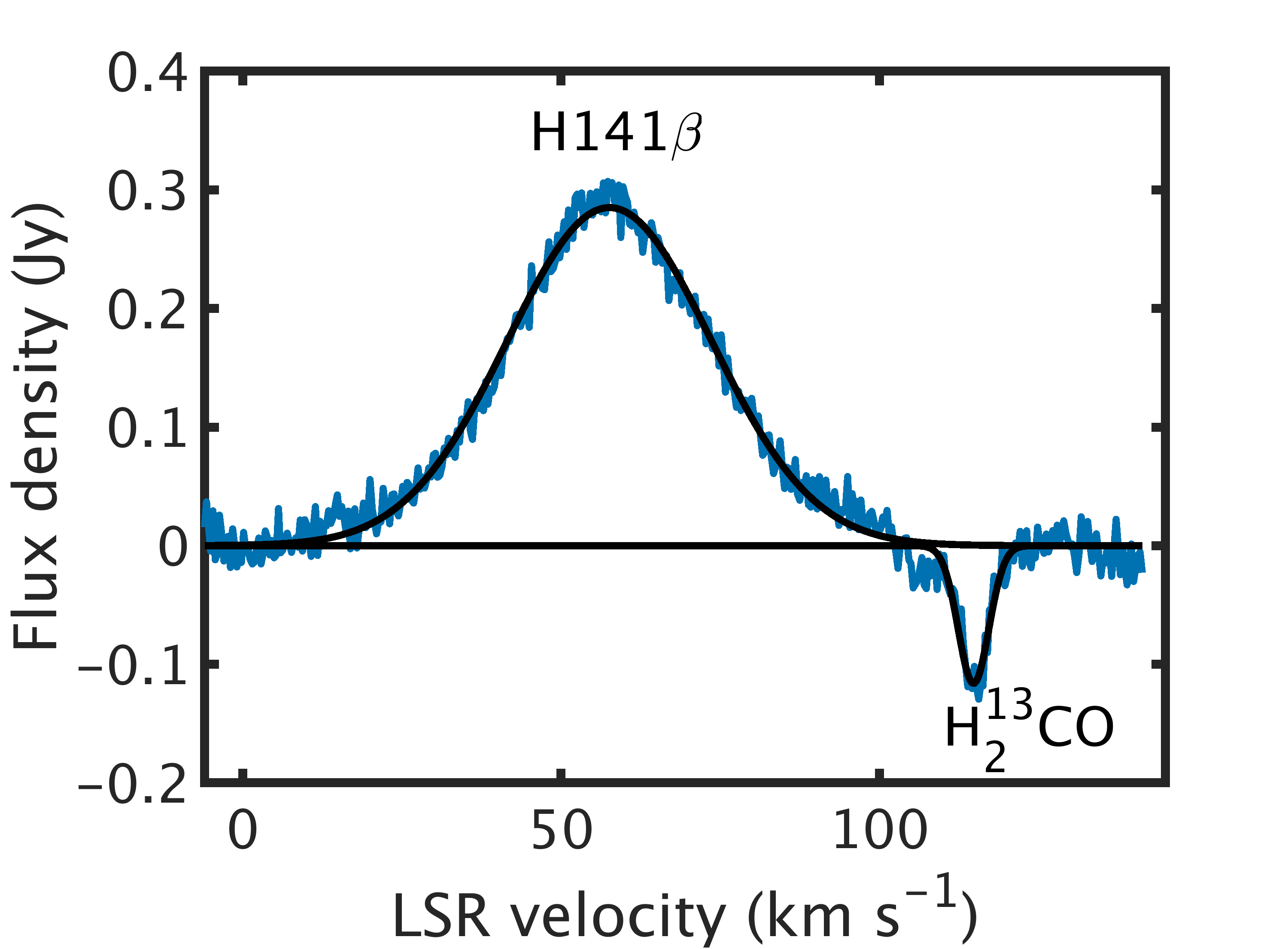}{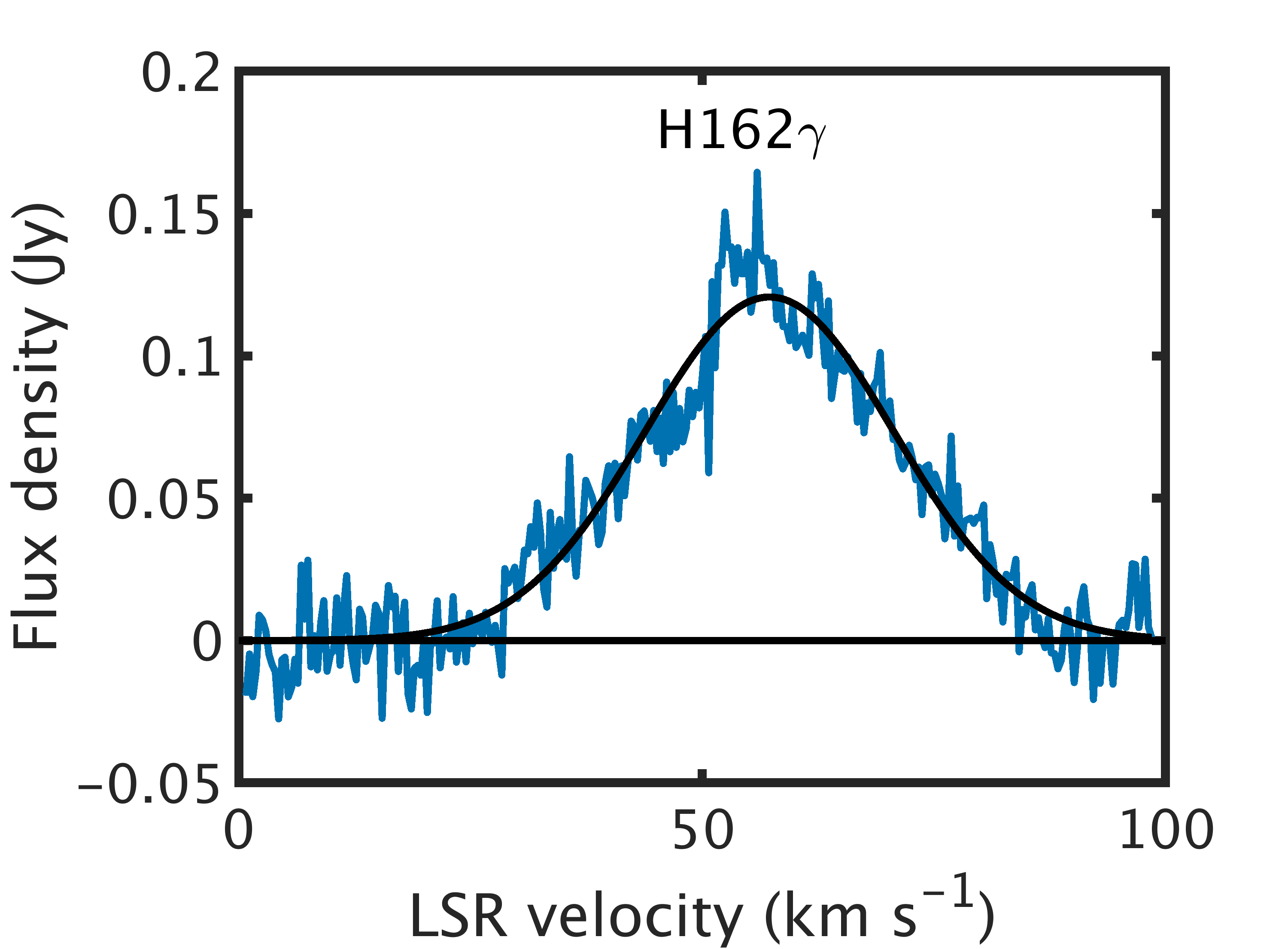}{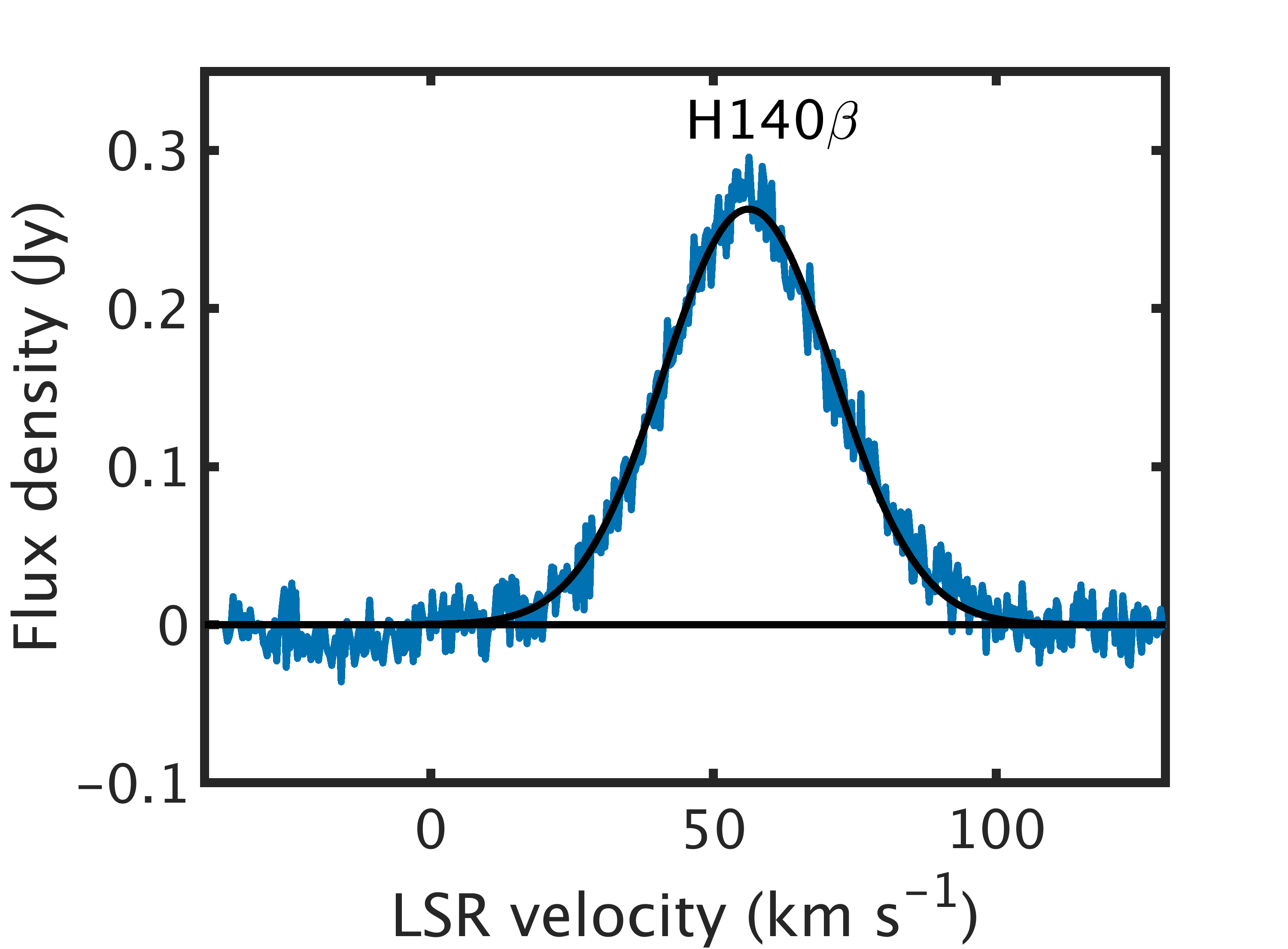}{fig4}{A few examples of the detected recombiation lines along with the Gaussian components (shown in black). The identified lines are marked on the spectra. The absorption feature on the top-right spectrum is due to a H$_2 ^{13}$CO transition.}

We have detected hydrogen, helium and carbon alpha transitions, and hydrogen beta and gamma transitions toward W51 IRS1 in the frequency range 4.5 to 5 GHz. A few molecular lines (such as Formaldehyde H$_2 ^{13}$CO and Formaldehyde H$_2$CO) were also detected in this frequency range; these results will be presented elsewhere. The transition, peak line flux density ($S_L$), central LSR velocity ($V_{LSR}$) and full width at half maximum ($\Delta V$) of the detected recombination lines are listed in Table~\ref{tab1}. The estimation errors given in Table~\ref{tab1} are formal errors obtained from Gaussian fitting (see below). A few example spectra (both recombination lines and, in one case, a molecular line) with Gaussian fits are shown in Fig.~\ref{fig4}. 

The mean LSR velocity of the source obtained from the hydrogen alpha transitions is 56.6$\pm$0.3 \kms. The LSR velocity obtained from the higher order transitions of hydrogen beta is 56.0$\pm$1.5 \kms. The uncertainty in the central velocity is larger than the statistical uncertainty in the LSR velocity obtained for each transition given in Table~\ref{tab1}, which we attribute to the residual baseline ripple affecting the line parameters. The LSR velocity obtained from the helium alpha transition is 56.4$\pm$1.2 \kms, comparable to that obtained from the hydrogen alpha transition.

The mean half power width of the hydrogen alpha line is 33.1$\pm$1.1 \kms. The line width of helium, estimated from a subset of spectra that are less affected by baseline ripple, is 28.3$\pm$0.3 \kms. The similarity of the central velocity of the hydrogen and helium lines indicates that
they originate from the same ionized gas. The thermal broadening of the helium line will be a factor of 2 smaller than that of hydrogen due to the
nuclear mass difference. We use this to estimate the electron temperature and turbulence line width, finding an electron temperature of 8500$\pm$1800 K and the turbulence half power width of 26.6$\pm$0.2 \kms.

The mean line flux density obtained from the same subset of data used for the electron temperature estimate
is 1.5$\pm$0.13 Jy. Previous observations indicate that the extent of the \HII\  region is $> 1$\arcmin\ \citep{md68,aetal79}. 
Assuming that the source is filling the telescope beam ($1^{'}\; \times\; 1^{'}$), the emission is close to LTE and the line optical depth at the observed frequency is $<<$ 1, the emission measure obtained is 5.4$\pm$2.7 $\times$ 10$^6$ pc cm$^{-6}$. This value is consistent with previous estimates \citep{m94}.

Our next step is to quantitatively assess the effect of spectral baseline ripple on the line parameters. It has been shown earlier that the spectral baseline ripple can be reduced by double position switching \citep{gs01}. We plan to re-process the 4.5-5 GHz data set using the double position switching method and compare the estimated line parameters with those presented in this paper. An improved spectral baseline will help us to investigate the presence of any weak, broad spectral lines toward W51A IRS1 that could possibly originate from the accretion of material by massive-star formation. Parameters of the broad line emission and the physical properties of the gas responsible for such emission, which could be well constrained from the wide-frequency observation, can put stringent constraints on star formation models.

\begin{table}[!ht]
\caption{Recombination Line parameters. 
\label{tab1}
}
\smallskip
\begin{center}
{\small
\begin{tabular}{lccc}  
\tableline
\noalign{\smallskip}
Transition & $S_L$ & $V_{LSR}$ & $\Delta V$\\
           & (Jy)  & (km s$^{-1}$) & (km s$^{-1}$) \\
\noalign{\smallskip}
\tableline
\noalign{\smallskip}
\multicolumn{4}{c}{$\alpha$ transitions}\\
\noalign{\smallskip}
\tableline
H109$\alpha$ & 1.610(0.002) &    56.52(0.02) &  31.4(0.1) \\ 
He109$\alpha$ & 0.255(0.003) &    54.42(0.37) &  30.9(0.7) \\ 
C109$\alpha^{a}$ & 0.098(0.005) &    54.78(0.65) &  20.9(1.2) \\ 
H110$\alpha$ & 1.520(0.002) &    56.51(0.02) &  31.2(0.1) \\ 
He110$\alpha$ & 0.192(0.003) &    56.80(0.22) &  29.9(0.6) \\ 
C110$\alpha$ & 0.053(0.004) &    57.04(0.47) &  11.8(1.1) \\ 
H111$\alpha$ & 1.596(0.004) &    56.22(0.04) &  34.4(0.1) \\ 
He111$\alpha$ & 0.152(0.005) &    56.38(0.48) &  28.7(1.3) \\ 
C111$\alpha$ & 0.050(0.009) &    57.56(0.78) &   9.1(1.9) \\ 
H112$\alpha$ & 1.389(0.003) &    57.03(0.03) &  32.6(0.1) \\
He112$\alpha$ & 0.118(0.003) &    57.56(0.60) &  28.1(1.4) \\ 
C112$\alpha^{a}$ & 0.103(0.004) &    55.45(0.55) &  19.9(1.2) \\ 
\tableline
\noalign{\smallskip}
\multicolumn{4}{c}{$\beta$ transitions}\\
\noalign{\smallskip}
\tableline
H137$\beta$ & 0.295(0.002) &    55.96(0.10) &  37.7(0.2) \\ 
H138$\beta$ & 0.245(0.002) &    57.05(0.11) &  33.3(0.2) \\ 
H139$\beta$ & 0.222(0.002) &    55.92(0.12) &  29.5(0.3) \\ 
H140$\beta$ &  0.263(0.002) &     56.23(0.12) &  35.5(0.3) \\ 
H141$\beta$ &  0.285(0.002) &     57.59(0.12) &  37.7(0.3) \\ 
\tableline
\noalign{\smallskip}
\multicolumn{4}{c}{$\gamma$ transitions}\\
\noalign{\smallskip}
\tableline
H157$\gamma$ &  0.101(0.002) &     56.92(0.27) &  31.8(0.6) \\ 
H161$\gamma$ &  0.106(0.002) &     53.09(0.31) &  38.7(0.7) \\ 
H162$\gamma$ &  0.121(0.002) &     57.26(0.25) &  31.7(0.6) \\
\noalign{\smallskip}
\tableline 
\multicolumn{4}{l}{$^a$ The line parameters of this transition are affected by} \\
\multicolumn{4}{l}{spectral baseline ripple}
\end{tabular}
}
\end{center}
\end{table}


\acknowledgements We thank Dr. Adam Ginsburg for providing the image of W51A, which was used to produce Fig. 1 in this paper. E.D.A. acknowledges partial support from NSF grant AST-1814063. The Arecibo Observatory is operated by the University of Central Florida under a cooperative agreement with the National Science Foundation (AST-1822073), and in alliance with Universidad Ana G. M\'{e}ndez and Yang Enterprises.

\bibliography{W51REUpaper}  

\end{document}